# COVID-19 epidemic outcome predictions based on logistic fitting and estimation of its reliability

Dávid Tátrai*, Zoltán Várallyay
Furukawa Electric Institute of Technology Ltd.
Késmárk str. 28A, 1158 Budapest, Hungary
* davidtatrai9@gmail.com

## Abstract

Since the first outbreak of the COVID-19 epidemic at the end of 2019, data has been made available on the number of infections, deaths and recoveries for all countries of the World, and that data can be used for statistical analysis. The primary interest of this paper is how well the logistic equation can predict the outcome of COVID-19 epidemic in any regions of the World assuming that the methodology of the testing process, namely the data collection method and social behavior is not changing over the course of time.

Besides the social relevance, this study has two scientific purposes: we investigate if a simple saturation model can describe the trend of the COVID-19 epidemic and if so, we would like to determine, from which point during the epidemic the fitting parameters provide reliable predictions. We also give estimations for the outcome of this epidemic in several countries based on the logistic model and the data available on 27 March, 2020. Based on the saturated cases in China, we have managed to find some criteria to judge the reliability of the predictions.

## Introduction

The outbreak of COVID-19 is the third time that a zoonotic coronavirus has crossed species to infect humans in the past two decades after SARS-CoV and MERS-CoV those with rather high lethality rate [1]. This type of corona virus appeared first and become an epidemic in Wuhan, Hupei, China and quickly spread over the World with an initially estimated reproductive number of 2.2 [2]. From data analysis point of view, the speed of spreading is decoded in the available data points even if the epidemic is still going on and not started to saturate yet. Of course, at the very beginning of the process the accuracy of the fitting and predictions can be questioned.

For a finalized epidemic, the curve for the cumulative number of infected cases mostly shows the behavior of the logistic growth as it is shown in Figure 1 which is the data for the Henan region, China. The cumulative number of infectious cases show such behavior that resembles to the well-known logistic growth, namely the outbreak has an exponentially increasing regime. This emerging segment is followed by an inflection point from that the number of daily infections starts to decrease. And finally, the cumulative number of cases is saturated, and no additional infection appears at the end of the process.

The idea to use this approach to describe and predict the epidemic growth comes from Ref. [3] where the number of deaths during the epidemic process is analyzed for Italy this way.

We similarly use the solution of the logistic equation in a different form (different mathematical basis) than in Ref. [3] to find the optimum parameters to the investigated dataset and calculate the expected duration and expected maximum number of infections. But in contrast we apply it for the number of registered infections instead of the number of deaths. Similar analysis, using a logistic fit was used in many additional works to predict the outcome of COVID-19 in selected regions recently [4-10]. We intend to extend this method to as much countries as possible and to make effort to describe the error of any prediction with the logistic model, and for that we introduce a variable to determine the reliability of the logistic fit. This parameter is evaluated based on the latest data point and the maximum number of cases given by the fitting. In this study, we analyze the trends of the spread of the COVID-19 epidemic by applying nonlinear least-square fitting method to get the best fit of the logistic curve to the available data. Our main purpose is trying to predict the evolution of the epidemic in several locations, but we note again that this work assumes that the testing methodology of the patients are not changing over time and the social behavior is also constant. For this analysis, we have used a publicly available database on the number of infected people by location [11].

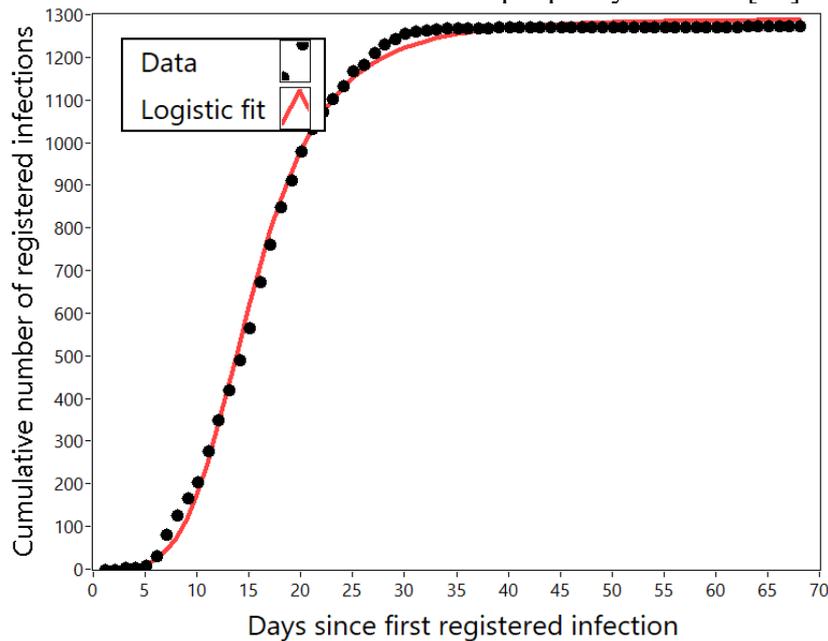

*Figure 1 Data from Henan region of China.*

## Methods

The logistic growth model or Verhulst model, after the name of Belgian mathematician Pierre Verhulst, is used to describe biological systems in connection with the population growth among different restrictions (limited resources for growing). We use the solution of the logistic growth equation [12] to fit its parameters to the investigated dataset. The solution can be given also in the following form among others [13]

$$N(t) = N_\infty - \frac{N_\infty - N_0}{1 + (t/t_c)^p} \qquad (1)$$

where $N$ is the number of infectious cases in this particular problem, $t$ is the time, $N_\infty$ is the cumulative number of infections at the end of the process or sometimes it is referred as the carrying capacity of the population, $N_0$ is the initial number of cases that we consider as 0 in every analyzed dataset, $t_c$ is the center of the curve and $p$ is the power parameter. From these variables, three of them have to be fitted in order to find an optimum fit to the particular dataset. These are the $N_\infty$, $t_c$ and $p$ parameters, respectively as a minimum number of parameters required for a saturation model.

The nonlinear fitting method what we use to obtain the best fit parameters is the Levenberg-Marquardt (LM) method that is a nonlinear least-square fitting method [14,15]. The results presented here were calculated by a custom-made code implemented in LabVIEW, by calling the LM fit routine.

Learning from the Chinese epidemic processes, we introduce some additional parameters we investigate throughout this paper. First, the cumulative number of registered infections (CNRI) which is the number of the positive tested cases since the outbreak, namely the measured value of $N(t)$ at the time $t$. Second, final value of the cumulative number of registered infections (FVCNRI) which is known only for saturated epidemics, like the ones in China. Ideally, this would be equal to $N_\infty$ in the model if the fitting was excellent. Third, the estimated final value of cumulative number of registered infections (EFVCNRI) which is a numerically fitted value on CNRI, namely $N_\infty$ in Eq. (1). We introduce the following ratio that we have found an essential parameter during the reliability investigations:

$$R = N_\infty / N(t_{now}) \qquad (2)$$

where $N_\infty$ is EFVCNRI while $N(t_{now})$ is CNRI, the latest measured data point we have.

This R value tells us where we are now in the logistic process if the fitted curve is well established. We have found, using ended epidemics from different regions of China that the logistic fitting gives different amount of error if we are at various parts of the logistic curve. The R value basically tells us the stability and reliability of the fitting and we have found that the prediction becomes reliable if $R<3$. We will discuss this in detail under the "Reliability of fit parameters" section.

## Results and discussions

For all countries and regions, we have fitted the logistic model. Based on the obtained parameters, we have calculated the estimated date for 50% infections ($N_\infty/2$) and the estimated date for 95% infections ($0.95 \cdot N_\infty$). For these, only the obtained parameters and the model description was used. The results including the R parameter are presented in Table 1 for all countries. We show an example for a non-saturated epidemic process; the data of Luxemburg is shown in Figure 2. The logistic model fits really well on the dataset and the inflection point of the epidemic is at day 27, according to the calculations.

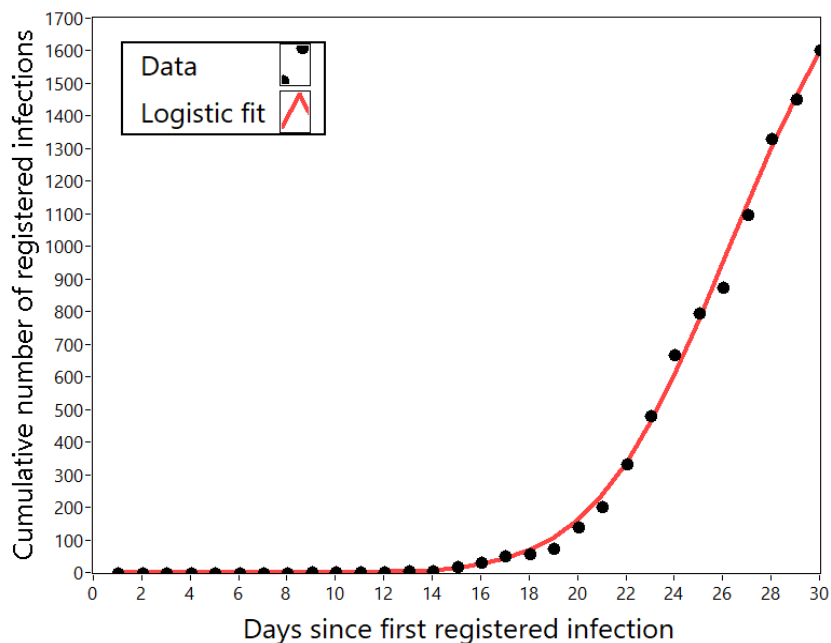

*Figure 2.: An example fit for a non-saturated epidemic. Data represents Luxemburg.*

## Reliability of fit parameters

Because the epidemic is in its first half in most locations, the reliability of the fitting parameters and so the derived parameters are questionable. Fortunately, for some regions in China, the epidemic seems to be saturated, so FCNVRI is known for those cases, and those datasets can be used to derive some reliability criteria for the found parameters of other locations too. For the saturated datasets, we have performed the logistic fit for the first 10; 11; 12… *M*-1; *M* day long data points (*M* = length of the dataset after the first registered infection) to estimate the behavior and convergence of the fit parameters. Now, we focus only on EFVCNRI, as the other parameters still need some additional investigations. We have observed two important facts:

- At the beginning of the dataset ($M \approx 10$), or a few days later, the EFVCNRI values usually 10-50 times bigger than FVCNRI, and from that time they start to converge smoothly (See Figure 3). Furthermore, from the extremely overestimated values, the converging time is relatively fast, it is in the order of 10-20 days. An example can be seen in Figure 4.
- A unified criterion has to be established to decide the reliability of the found parameters without knowing FVCNRI. We have previously introduced the *R* parameter, which can be calculated from the fit EFVCNRI ($N_\infty$) and the most recent CNRI ($N(t_{now})$) values. We have calculated this R parameter from the day number 10 to all available days in the datasets for China. We have found that the R parameter converges to 1 (the ideal value at the end of the epidemic) similarly as EFVCNRI converges to FVCNRI. And the two convergence curves have good correlation. Examples for Hainan and Heilongjiang regions of China can be found in Figure 3.

After investigating all saturated datasets from China, we have found the following set of criteria to judge the results:

1. If the dataset is over the "huge overestimation" region and 2<R< 3, then the EFVCNRI will be overestimated by not more than a factor of 2.
2. If the dataset is over the "huge overestimation" region and R=2, (namely, the fit shows that the pandemic is at 50% considering CNRI) then the EFVCNRI will be overestimated by a factor of 1.3-1.7. That means that the epidemic is closer to the end than estimated.
3. If the dataset is over the "huge overestimation" region and R<2, (namely, the fit shows that the pandemic is over 50%) then the fit parameters will be close to the real ones with moderate error.
4. And if R>>3 the reliability of the fit parameters will be very poor, but generally we can state that EFVCNRI is overestimated, and the overestimation factor can be 5-30 or even more.

Based on these observations, we have calculated the R parameter for all countries and regions and included them in Table 1. Furthermore, in all cases, where R>3 we have highlighted the corresponding row with red to emphasize the non-reliability of those data. In Table 1, pure calculation results are presented, and no any correction based on the R value was applied.

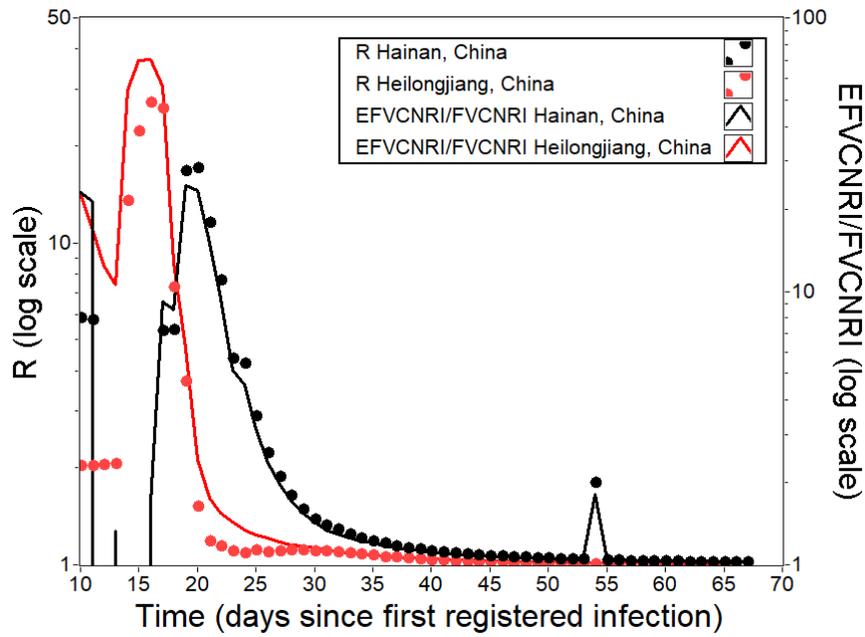

*Figure 3.: The convergence of R parameter and EFVCNRI/FVCNRI ratio over time for Hainan and Heilongjiang regions of China.*

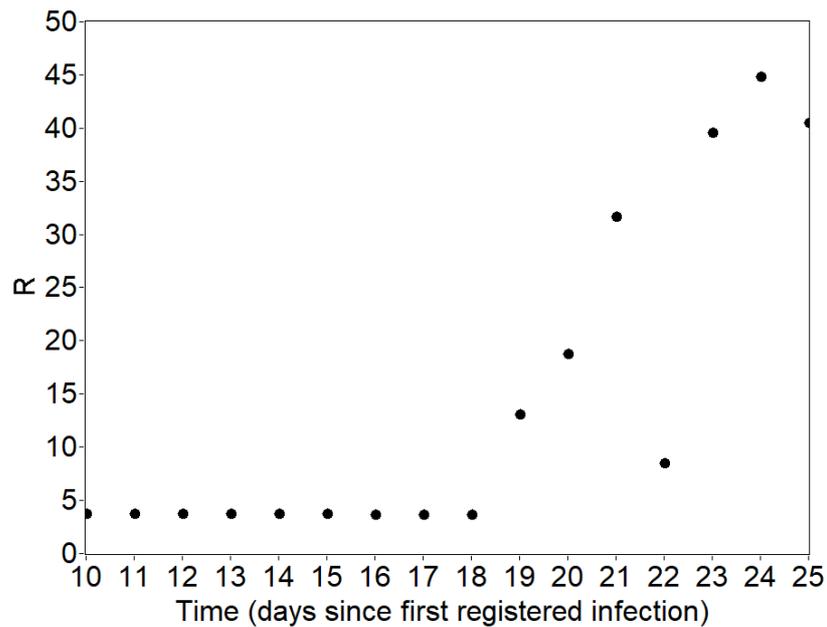

*Figure 4.: An example for the increase of the R parameter over time for a non-saturated epidemic. It is important to emphasize that such behavior seems to be a standard feature. Furthermore, this increase of the R parameter is an indicator that the dataset and the prediction of the logistic model is expected to become reliable soon. Data is for Hungary.*

## Limitations

While fitting the logistic model on the datasets for the various regions and countries listed in the dataset, many cases, the number of registered infections was too few or too "noisy" to be able to make any realistic fit on them. The data for those places is not published in this study and denoted with NaN.

Some cases, the trend of the epidemic does not follow a logistic behavior such as in cases of Figure 5 and Figure 6.

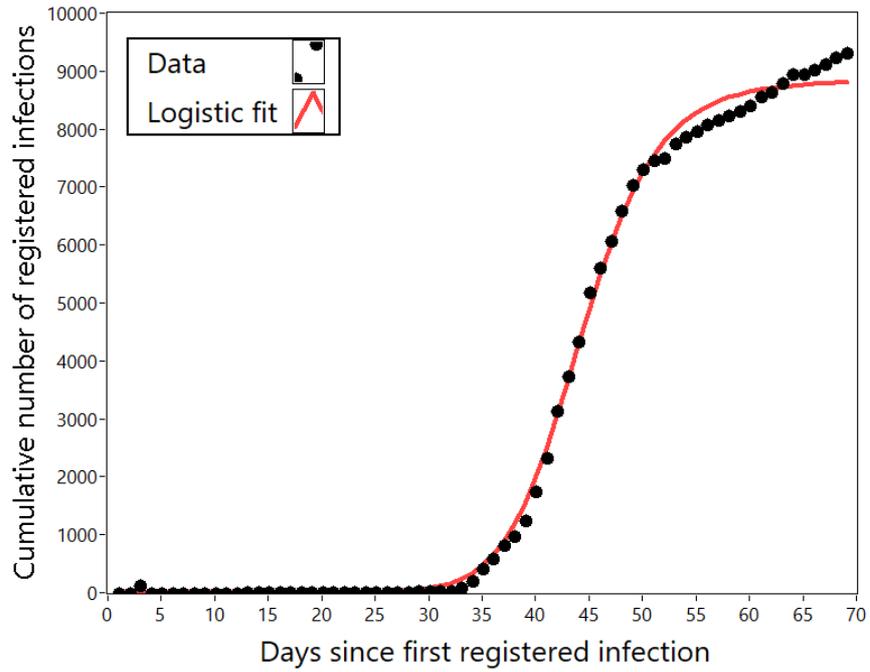

*Figure 5.: The data for Korea. Around day 50, the trend has broken and significantly deviates from logistic behavior. Changes of the methodology of registration of infections or severe political restriction measures can cause such breaks of the trends.*

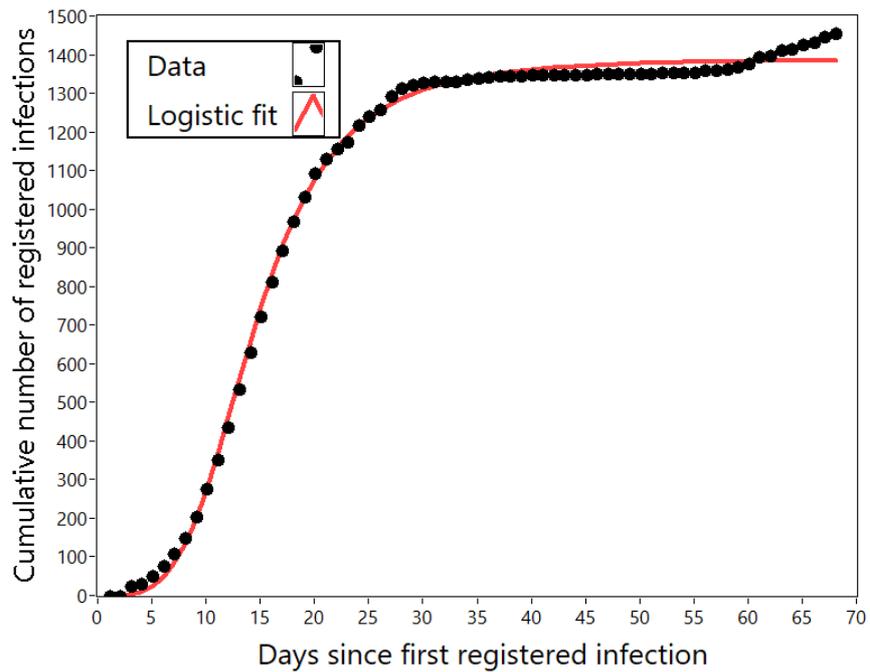

*Figure 6.: A starting second wave of the epidemic around day 60 at Guangdong, China.*

Unfortunately, in some regions of China (Figure 6), after a saturated period, a second wave of the epidemic can be observed. Those regions will be studied later because such second waves can happen in the future in any other locations too.

In all these kinds of cases, we provide NaN (Not a Number) as a result in Table 1.

| Location | | Estimated date for 50% of the epidemic | Estimated date for 95% of the epidemic | EFVCNRI | R |
|---|---|---|---|---|---|
| Region | Country | | | | |
| | Afghanistan | 2020.04.09 | 2020.05.09 | 972 | 8,84 |
| | Albania | 2020.12.05 | 2020.12.30 | 28081 | 150,97 |
| | Algeria | 2020.04.04 | 2020.05.01 | 1897 | 4,64 |
| | Andorra | 2020.03.27 | 2020.04.14 | 576 | 2,16 |
| | Angola | 2022.09.13 | 2021.01.10 | 637 | 159,15 |
| | Antigua and Barbuda | 2020.05.06 | 2020.07.07 | 974 | 139,12 |
| | Argentina | 2020.04.04 | 2020.05.02 | 2904 | 4,93 |
| | Armenia | 2020.03.22 | 2020.04.06 | 421 | 1,28 |
| Australian Capital Territory | Australia | 2020.04.14 | 2020.05.28 | 981 | 15,83 |
| New South Wales | Australia | 2020.04.12 | 2020.05.06 | 21188 | 15,08 |
| Northern Territory | Australia | 2020.05.03 | 2020.07.03 | 896 | 74,69 |
| Queensland | Australia | 2020.03.26 | 2020.04.10 | 1023 | 1,84 |
| South Australia | Australia | 2020.03.30 | 2020.04.14 | 800 | 3,11 |
| Tasmania | Australia | 2020.04.18 | 2020.06.06 | 731 | 15,55 |
| Victoria | Australia | 2020.03.24 | 2020.04.05 | 837 | 1,46 |
| Western Australia | Australia | 2020.03.25 | 2020.04.10 | 412 | 1,78 |
| | Austria | 2020.04.02 | 2020.04.29 | 28880 | 3,77 |
| | Azerbaijan | 2020.04.21 | 2020.06.15 | 2074 | 12,57 |
| | Bahamas | 2020.08.29 | 2021.01.06 | 988 | 98,78 |
| | Bahrain | 2020.07.22 | 2020.12.16 | 8798 | 18,88 |
| | Bangladesh | 2020.05.10 | 2020.10.08 | 856 | 17,84 |
| | Barbados | 2020.07.16 | 2021.01.07 | 980 | 40,85 |
| | Belarus | 2020.03.18 | 2020.04.06 | 113 | 1,21 |
| | Belgium | 2020.04.19 | 2020.05.21 | 172684 | 23,71 |
| | Benin | 2021.03.04 | 2021.01.06 | 955 | 159,1 |
| | Bhutan | 2023.07.13 | 2020.12.27 | 63 | 20,84 |
| | Bolivia | 2020.05.22 | 2020.11.10 | 1541 | 25,26 |
| | Bosnia and Herzegovina | 2020.03.27 | 2020.04.20 | 478 | 2,02 |
| | Brazil | 2020.03.25 | 2020.04.06 | 5288 | 1,55 |
| | Brunei | NaN | NaN | NaN | NaN |
| | Bulgaria | 2020.03.26 | 2020.04.29 | 573 | 1,96 |
| | Burkina Faso | 2020.03.26 | 2020.04.12 | 325 | 1,8 |
| | Cabo Verde | 2022.10.21 | 2021.01.10 | 570 | 113,94 |
| | Cambodia | 2020.03.20 | 2020.03.28 | 107 | 1,08 |

*Table 1 part 1.: The estimated parameters for the various locations. NaN represents cases, where the curve fitting did not converge well, or the available data contained only very few points, or other cases described in the limitations section. The R parameter is also presented for all locations. If the R parameter is greater than 3 than the estimated parameters are not reliable, the EFVCNRI value is probably highly overestimated, so those lines are highlighted in red.*

| Location | | Estimated date for 50% of the epidemic | Estimated date for 95% of the epidemic | EFVCNRI | R |
|---|---|---|---|---|---|
| Region | Country | | | | |
|  | Cameroon | 2020.03.23 | 2020.04.02 | 111 | 1,22 |
| Alberta | Canada | 2020.04.08 | 2020.05.19 | 3279 | 6,05 |
| British Columbia | Canada | 2020.03.22 | 2020.04.03 | 971 | 1,34 |
| Grand Princess | Canada | 2021.12.24 | 2021.01.03 | 725 | 55,78 |
| Manitoba | Canada | 2020.07.06 | 2021.01.03 | 962 | 24,66 |
| New Brunswick | Canada | 2020.05.10 | 2020.09.17 | 967 | 21,49 |
| Newfoundland and Labrador | Canada | 2020.04.03 | 2020.04.18 | 1034 | 10,14 |
| Nova Scotia | Canada | 2020.04.17 | 2020.07.07 | 995 | 11,05 |
| Ontario | Canada | 2020.04.16 | 2020.05.15 | 19076 | 19,19 |
| Prince Edward Island | Canada | 2020.06.28 | 2021.01.05 | 987 | 109,7 |
| Quebec | Canada | 2020.03.24 | 2020.03.31 | 2585 | 1,28 |
| Saskatchewan | Canada | 2020.03.22 | 2020.04.01 | 116 | 1,22 |
|  | Central African Republic | 2023.08.04 | 2021.01.05 | 213 | 70,94 |
|  | Chad | 2021.10.08 | 2021.01.09 | 898 | 299,37 |
|  | Chile | 2020.04.12 | 2020.05.20 | 18277 | 11,35 |
| Anhui | China | 2020.02.03 | 2020.02.19 | 1004 | 1,01 |
| Beijing | China | 2020.02.06 | 2020.04.19 | 534 | 0,94 |
| Chongqing | China | 2020.02.01 | 2020.02.22 | 588 | 1,02 |
| Fujian | China | 2020.02.01 | 2020.02.22 | 307 | 0,93 |
| Gansu | China | 2020.02.18 | 2020.10.16 | 180 | 1,32 |
| Guangdong | China | 2020.02.02 | 2020.02.20 | 1389 | 0,95 |
| Guangxi | China | 2020.02.02 | 2020.02.25 | 259 | 1,02 |
| Guizhou | China | 2020.02.05 | 2020.02.19 | 148 | 1,01 |
| Hainan | China | 2020.02.03 | 2020.02.23 | 172 | 1,03 |
| Hebei | China | 2020.02.05 | 2020.02.26 | 326 | 1,02 |
| Heilongjiang | China | 2020.02.06 | 2020.02.22 | 488 | 1,01 |
| Henan | China | 2020.02.03 | 2020.02.19 | 1289 | 1,01 |
| Hong Kong | China | 2020.10.16 | 2020.11.14 | 84337 | 162,5 |
| Hubei | China | NaN | NaN | NaN | NaN |
| Hunan | China | 2020.02.02 | 2020.02.18 | 1031 | 1,01 |
| Inner Mongolia | China | 2020.02.04 | 2020.03.01 | 79 | 0,86 |
| Jiangsu | China | 2020.02.04 | 2020.02.21 | 644 | 1 |
| Jiangxi | China | 2020.02.03 | 2020.02.18 | 946 | 1,01 |
| Jilin | China | NaN | NaN | NaN | NaN |
| Liaoning | China | NaN | NaN | NaN | NaN |
| Macau | China | 2043.08.06 | 2020.11.13 | 897 | 27,18 |

*Table 1 part 2.: The estimated parameters for the various locations. NaN represents cases, where the curve fitting did not converge well, or the available data contained only very few points, or other cases described in the limitations section. The R parameter is also presented for all locations. If the R parameter is greater than 3 than the estimated parameters are not reliable, the EFVCNRI value is probably highly overestimated, so those lines are highlighted in red.*

| Location | | Estimated date for 50% of the epidemic | Estimated date for 95% of the epidemic | EFVCNRI | R |
|---|---|---|---|---|---|
| Region | Country | | | | |
| Ningxia | China | 2020.02.04 | 2020.03.01 | 77 | 1,03 |
| Qinghai | China | NaN | NaN | NaN | NaN |
| Shaanxi | China | NaN | NaN | NaN | NaN |
| Shandong | China | 2020.02.07 | 2020.03.17 | 818 | 1,06 |
| Shanghai | China | 2020.02.02 | 2020.03.01 | 376 | 0,8 |
| Shanxi | China | 2020.02.02 | 2020.02.17 | 135 | 1 |
| Sichuan | China | 2020.02.03 | 2020.02.29 | 556 | 1,01 |
| Tianjin | China | 2020.02.04 | 2020.03.01 | 143 | 0,92 |
| Tibet | China | 2020.10.13 | 2020.11.21 | 2 | 2,48 |
| Xinjiang | China | NaN | NaN | NaN | NaN |
| Yunnan | China | 2020.01.31 | 2020.02.19 | 177 | 0,98 |
| Zhejiang | China | 2020.02.01 | 2020.02.16 | 1229 | 0,99 |
| | Colombia | 2020.03.26 | 2020.04.13 | 1091 | 2,02 |
| | Congo (Brazzaville) | 2023.01.30 | 2021.01.05 | 358 | 89,46 |
| | Congo (Kinshasa) | 2020.03.21 | 2020.03.30 | 59 | 1,16 |
| | Costa Rica | 2020.04.14 | 2020.06.12 | 1980 | 7,53 |
| | Cote d'Ivoire | 2020.03.23 | 2020.03.27 | 106 | 1,04 |
| | Croatia | 2020.03.28 | 2020.04.14 | 1337 | 2,28 |
| | Diamond Princess | 2020.02.15 | 2020.02.26 | 718 | 1,01 |
| | Cuba | 2020.04.16 | 2020.06.13 | 963 | 12,04 |
| | Cyprus | 2020.04.02 | 2020.06.02 | 477 | 2,94 |
| | Czechia | 2020.04.11 | 2020.05.26 | 15482 | 6,79 |
| Faroe Islands | Denmark | 2020.03.19 | 2020.03.29 | 152 | 1,05 |
| Greenland | Denmark | 2020.07.16 | 2021.01.06 | 990 | 98,99 |
| | Denmark | 2020.03.20 | 2020.04.20 | 2642 | 1,29 |
| | Djibouti | 2020.04.28 | 2020.07.02 | 997 | 83,07 |
| | Dominican Republic | 2020.03.25 | 2020.04.05 | 945 | 1,63 |
| | Ecuador | 2020.03.22 | 2020.03.31 | 1799 | 1,13 |
| | Egypt | 2020.03.27 | 2020.04.25 | 1119 | 2,09 |
| | El Salvador | 2020.05.11 | 2020.09.06 | 997 | 76,73 |
| | Equatorial Guinea | 2021.01.22 | 2021.01.05 | 952 | 79,32 |

*Table 1 part 3.: The estimated parameters for the various locations. NaN represents cases, where the curve fitting did not converge well, or the available data contained only very few points, or other cases described in the limitations section. The R parameter is also presented for all locations. If the R parameter is greater than 3 than the estimated parameters are not reliable, the EFVCNRI value is probably highly overestimated, so those lines are highlighted in red.*

| Location | | Estimated date for 50% of the epidemic | Estimated date for 95% of the epidemic | EFVCNRI | R |
| --- | --- | --- | --- | --- | --- |
| Region | Country | | | | |
| | Eritrea | 2020.05.16 | 2020.09.14 | 999 | 166,5 |
| | Estonia | 2020.03.26 | 2020.04.27 | 986 | 1,71 |
| | Eswatini | 2020.05.31 | 2020.10.19 | 982 | 109,06 |
| | Ethiopia | 2021.03.10 | 2021.01.03 | 908 | 56,72 |
| | Fiji | 2021.01.02 | 2021.01.09 | 970 | 194,05 |
| | Finland | 2020.03.25 | 2020.04.14 | 1881 | 1,81 |
| French Guiana | France | 2020.08.10 | 2020.12.28 | 868 | 30,99 |
| French Polynesia | France | 2020.05.29 | 2021.01.03 | 958 | 31,93 |
| Guadeloupe | France | NaN | NaN | NaN | NaN |
| Mayotte | France | 2020.04.20 | 2020.06.25 | 985 | 19,69 |
| New Caledonia | France | 2020.07.03 | 2021.01.09 | 992 | 66,15 |
| Reunion | France | 2020.03.25 | 2020.04.09 | 246 | 1,7 |
| Saint Barthelemy | France | 2023.04.02 | 2020.12.25 | 22 | 4,49 |
| St Martin | France | 2020.07.29 | 2020.12.30 | 913 | 83,04 |
| Martinique | France | 2020.04.22 | 2020.07.20 | 836 | 10,33 |
| | France | 2020.04.02 | 2020.04.25 | 116741 | 3,54 |
| | Gabon | 2020.11.25 | 2021.01.04 | 962 | 137,41 |
| | Gambia | 2022.03.28 | 2021.01.07 | 820 | 273,46 |
| | Georgia | 2020.06.08 | 2020.12.18 | 1452 | 17,5 |
| | Germany | 2020.03.28 | 2020.04.15 | 111347 | 2,19 |
| | Ghana | 2020.03.25 | 2020.04.02 | 222 | 1,62 |
| | Greece | 2020.04.02 | 2020.05.22 | 2734 | 2,83 |
| | Guatemala | 2020.07.21 | 2021.01.04 | 954 | 34,08 |
| | Guinea | 2020.06.22 | 2021.01.03 | 973 | 121,63 |
| | Guyana | 2022.12.28 | 2021.01.02 | 227 | 11,36 |
| | Haiti | 2021.01.06 | 2021.01.10 | 963 | 120,42 |
| | Holy See | 2020.08.17 | 2020.12.27 | 844 | 211,09 |
| | Honduras | 2020.04.24 | 2020.07.15 | 967 | 14,22 |
| | Hungary | 2020.05.02 | 2020.07.14 | 9107 | 30,36 |
| | Iceland | 2020.04.14 | 2020.06.17 | 5672 | 6,37 |
| | India | 2020.04.16 | 2020.05.16 | 16493 | 18,59 |
| | Indonesia | 2020.04.05 | 2020.05.12 | 4498 | 4,3 |
| | Iran | NaN | NaN | NaN | NaN |
| | Iraq | 2020.04.25 | 2020.08.04 | 2997 | 6,54 |
| | Ireland | 2020.03.30 | 2020.04.21 | 5785 | 2,73 |
| | Israel | 2020.04.15 | 2020.05.08 | 116428 | 38,36 |

*Table 1 part 4.: The estimated parameters for the various locations. NaN represents cases, where the curve fitting did not converge well, or the available data contained only very few points, or other cases described in the limitations section. The R parameter is also presented for all locations. If the R parameter is greater than 3 than the estimated parameters are not reliable, the EFVCNRI value is probably highly overestimated, so those lines are highlighted in red.*

| Location | | Estimated date for 50% of the epidemic | Estimated date for 95% of the epidemic | EFVCNRI | R |
|---|---|---|---|---|---|
| Region | Country | | | | |
|  | Italy | NaN | NaN | NaN | NaN |
|  | Jamaica | 2021.02.20 | 2021.01.01 | 873 | 33,58 |
|  | Japan | 2020.04.17 | 2020.07.28 | 5497 | 3,74 |
|  | Jordan | 2020.03.27 | 2020.04.18 | 506 | 2,15 |
|  | Kazakhstan | 2020.08.23 | 2021.01.03 | 14577 | 97,18 |
|  | Kenya | 2020.05.06 | 2020.08.31 | 964 | 31,09 |
|  | Korea | NaN | NaN | NaN | NaN |
|  | Kuwait | NaN | NaN | NaN | NaN |
|  | Kyrgyzstan | NaN | NaN | NaN | NaN |
|  | Latvia | 2020.03.27 | 2020.04.21 | 590 | 2,11 |
|  | Lebanon | 2020.05.10 | 2020.08.12 | 8356 | 21,37 |
|  | Liberia | NaN | NaN | NaN | NaN |
|  | Liechtenstein | 2020.03.19 | 2020.03.28 | 60 | 1,07 |
|  | Lithuania | 2020.03.24 | 2020.04.03 | 506 | 1,41 |
|  | Luxembourg | 2020.03.23 | 2020.04.04 | 2226 | 1,39 |
|  | Madagascar | NaN | NaN | NaN | NaN |
|  | Malaysia | 2020.03.23 | 2020.04.10 | 3268 | 1,51 |
|  | Maldives | NaN | NaN | NaN | NaN |
|  | Malta | 2020.03.25 | 2020.04.21 | 252 | 1,82 |
|  | Mauritania | NaN | NaN | NaN | NaN |
|  | Mauritius | 2020.04.11 | 2020.06.04 | 1003 | 10,67 |
|  | Mexico | 2020.03.30 | 2020.04.22 | 1735 | 2,97 |
|  | Moldova | 2020.04.06 | 2020.05.20 | 959 | 4,82 |
|  | Monaco | 2020.04.18 | 2020.06.01 | 656 | 15,61 |
|  | Mongolia | NaN | NaN | NaN | NaN |
|  | Montenegro | 2020.04.13 | 2020.06.07 | 998 | 12,18 |
|  | Morocco | 2020.04.08 | 2020.05.08 | 2983 | 8,65 |
|  | Namibia | NaN | NaN | NaN | NaN |
|  | Nepal | NaN | NaN | NaN | NaN |
| Aruba | Netherlands | 2020.04.16 | 2020.05.25 | 978 | 29,62 |
| Curacao | Netherlands | NaN | NaN | NaN | NaN |
| Sint Maarten | Netherlands | NaN | NaN | NaN | NaN |
|  | Netherlands | 2020.04.28 | 2020.07.01 | 190602 | 22,16 |
|  | New Zealand | 2020.04.01 | 2020.04.15 | 1904 | 5,17 |
|  | Nicaragua | NaN | NaN | NaN | NaN |
|  | Niger | NaN | NaN | NaN | NaN |
|  | Nigeria | 2020.04.07 | 2020.05.04 | 653 | 9,33 |

*Table 1 part 5.: The estimated parameters for the various locations. NaN represents cases, where the curve fitting did not converge well, or the available data contained only very few points, or other cases described in the limitations section. The R parameter is also presented for all locations. If the R parameter is greater than 3 than the estimated parameters are not reliable, the EFVCNRI value is probably highly overestimated, so those lines are highlighted in red.*

| Location | | Estimated date for 50% of the epidemic | Estimated date for 95% of the epidemic | EFVCNRI | R |
|---|---|---|---|---|---|
| Region | Country | | | | |
| | North Macedonia | 2020.03.24 | 2020.04.07 | 339 | 1,55 |
| | Norway | 2020.04.06 | 2020.06.09 | 12326 | 3,28 |
| | Oman | NaN | NaN | NaN | NaN |
| | Pakistan | 2020.03.22 | 2020.04.03 | 1650 | 1,2 |
| | Panama | 2020.04.22 | 2020.06.27 | 11910 | 17,67 |
| | Papua New Guinea | NaN | NaN | NaN | NaN |
| | Paraguay | 2020.05.07 | 2020.08.18 | 1310 | 25,18 |
| | Peru | 2020.03.26 | 2020.04.20 | 1187 | 1,87 |
| | Philippines | 2020.04.04 | 2020.04.29 | 4033 | 5,02 |
| | Poland | 2020.04.10 | 2020.05.23 | 9712 | 6,99 |
| | Portugal | 2020.04.07 | 2020.05.05 | 30259 | 7,09 |
| | Qatar | NaN | NaN | NaN | NaN |
| | Romania | 2020.04.24 | 2020.06.05 | 40359 | 31,24 |
| | Russia | 2020.04.15 | 2020.05.08 | 31744 | 30,64 |
| | Rwanda | NaN | NaN | NaN | NaN |
| | Saint Lucia | NaN | NaN | NaN | NaN |
| | Saint Vincent and the Grenadines | NaN | NaN | NaN | NaN |
| | San Marino | 2020.04.04 | 2020.07.07 | 628 | 2,82 |
| | Saudi Arabia | 2020.04.02 | 2020.04.30 | 4256 | 3,85 |
| | Senegal | 2020.04.10 | 2020.05.29 | 723 | 6,08 |
| | Serbia | 2020.04.05 | 2020.05.14 | 2125 | 4,65 |
| | Seychelles | NaN | NaN | NaN | NaN |
| | Singapore | NaN | NaN | NaN | NaN |
| | Slovakia | 2020.03.21 | 2020.04.13 | 350 | 1,3 |
| | Slovenia | 2020.03.30 | 2020.06.04 | 1400 | 2,22 |
| | Somalia | NaN | NaN | NaN | NaN |
| | South Africa | 2020.04.18 | 2020.05.19 | 47467 | 40,57 |
| | Spain | 2020.04.04 | 2020.04.27 | 340216 | 5,18 |
| | Sri Lanka | 2020.03.18 | 2020.03.26 | 110 | 1,04 |
| | Sudan | NaN | NaN | NaN | NaN |
| | Suriname | NaN | NaN | NaN | NaN |
| | Sweden | 2020.03.26 | 2020.04.24 | 5818 | 1,9 |
| | Switzerland | 2020.03.25 | 2020.04.13 | 22283 | 1,72 |
| | Taiwan* | NaN | NaN | NaN | NaN |
| | Tanzania | NaN | NaN | NaN | NaN |

*Table 1 part 6.: The estimated parameters for the various locations. NaN represents cases, where the curve fitting did not converge well, or the available data contained only very few points, or other cases described in the limitations section. The R parameter is also presented for all locations. If the R parameter is greater than 3 than the estimated parameters are not reliable, the EFVCNRI value is probably highly overestimated, so those lines are highlighted in red.*

| Location | | Estimated date for 50% of the epidemic | Estimated date for 95% of the epidemic | EFVCNRI | R |
|---|---|---|---|---|---|
| Region | Country | | | | |
| | Thailand | 2020.03.25 | 2020.04.07 | 1929 | 1,7 |
| | Togo | NaN | NaN | NaN | NaN |
| | Trinidad and Tobago | NaN | NaN | NaN | NaN |
| | Tunisia | 2020.04.20 | 2020.06.01 | 6646 | 29,28 |
| | Turkey | NaN | NaN | NaN | NaN |
| | Uganda | NaN | NaN | NaN | NaN |
| | Ukraine | 2020.04.18 | 2020.05.10 | 48981 | 158 |
| | United Arab Emirates | NaN | NaN | NaN | NaN |
| Bermuda | United Kingdom | NaN | NaN | NaN | NaN |
| Cayman Islands | United Kingdom | NaN | NaN | NaN | NaN |
| Channel Islands | United Kingdom | 2020.04.12 | 2020.05.23 | 951 | 10,8 |
| Gibraltar | United Kingdom | NaN | NaN | NaN | NaN |
| Isle of Man | United Kingdom | NaN | NaN | NaN | NaN |
| Montserrat | United Kingdom | NaN | NaN | NaN | NaN |
| | United Kingdom | 2020.04.14 | 2020.05.09 | 282643 | 19,43 |
| | Uruguay | 2020.03.22 | 2020.04.10 | 312 | 1,31 |
| | US | 2020.03.28 | 2020.04.10 | 272177 | 2,68 |
| | Uzbekistan | NaN | NaN | NaN | NaN |
| | Venezuela | 2020.03.25 | 2020.05.05 | 191 | 1,78 |
| | Vietnam | 2020.05.13 | 2020.08.08 | 2930 | 17,98 |
| | Zambia | NaN | NaN | NaN | NaN |
| | Zimbabwe | NaN | NaN | NaN | NaN |
| Diamond Princess | Canada | NaN | NaN | NaN | NaN |
| | Dominica | NaN | NaN | NaN | NaN |
| | Grenada | NaN | NaN | NaN | NaN |
| | Mozambique | NaN | NaN | NaN | NaN |
| | Syria | NaN | NaN | NaN | NaN |
| | Timor-Leste | NaN | NaN | NaN | NaN |
| | Belize | NaN | NaN | NaN | NaN |
| Recovered | Canada | NaN | NaN | NaN | NaN |
| | Laos | NaN | NaN | NaN | NaN |
| | Libya | NaN | NaN | NaN | NaN |
| | West Bank and Gaza | NaN | NaN | NaN | NaN |
| | Guinea-Bissau | NaN | NaN | NaN | NaN |
| | Mali | NaN | NaN | NaN | NaN |

*Table 1 part 7.: The estimated parameters for the various locations. NaN represents cases, where the curve fitting did not converge well, or the available data contained only very few points, or other cases described in the limitations section. The R parameter is also presented for all locations. If the R parameter is greater than 3 than the estimated parameters are not reliable, the EFVCNRI value is probably highly overestimated, so those lines are highlighted in red.*

| Location | | Estimated date for 50% of the epidemic | Estimated date for 95% of the epidemic | EFVCNRI | R |
|---|---|---|---|---|---|
| Region | Country | | | | |
|  | Saint Kitts and Nevis | NaN | NaN | NaN | NaN |
| Northwest Territories | Canada | NaN | NaN | NaN | NaN |
| Yukon | Canada | NaN | NaN | NaN | NaN |
|  | Kosovo | NaN | NaN | NaN | NaN |
|  | Burma | NaN | NaN | NaN | NaN |

*Table 1 part 8.: The estimated parameters for the various locations. NaN represents cases, where the curve fitting did not converge well, or the available data contained only very few points, or other cases described in the limitations section. The R parameter is also presented for all locations. If the R parameter is greater than 3 than the estimated parameters are not reliable, the EFVCNRI value is probably highly overestimated, so those lines are highlighted in red.*

## The continuation of the work

As, the world is in the middle or in the first half of the epidemic, the authors are planning to continue their work, possibly with weekly updates and reanalysis of the updated dataset. Based on the present fit results, several countries are around the inflection point of the epidemic, and consequently, the EFVCNRI is already in the converging period with R~2 values. During the next week, EFVCNRI should continue converging, and in those regions, signs of the saturation might be observable. Furthermore, we expect that countries with R>10 values are also getting within the converging regime soon to predict the outcome of their epidemic process more accurately.

## Acknowledgements

The authors would like to express their appreciation for everybody fighting against the COVID-19 pandemic.